\documentstyle[preprint,aps,floats]{revtex}

\input epsf.tex
\def\DESepsf(#1 width #2){\epsfxsize=#2 \epsfbox{#1}}

\newcommand{\Jnl}[4]{#1 {\bf #2} (#4) #3}

\begin{document}
\preprint{\vbox{\hbox{UM-P-016-2002}}}
\title{Magnetic Properties of Scalar Particles --- the
Scalar Aharonov-Casher Effect and Supersymmetry  }

\author{Xiao-Gang He$^{1,2}$ and Bruce H. McKellar$^3$}
\address{
$^1$ Department of Physics, Peking University, Beijing 100871\\
$^2$Department of Physics, National Taiwan University,
Taipei\\
$^3$ School of Physics, University of Melbourne, Parkville,
Vic 3052}

\maketitle
\begin{abstract}

The original topological Aharonov-Casher (AC) effect is due to the
interaction of the anomalous magnetic dipole moment (MDM) with certain
configurations of electric field.  Naively one would not expect an AC
effect for a scalar particle for which no anomalous MDM can be defined
in the usual sense.  In this letter we study the AC effect in
supersymmetric systems.  In this framework there is the possibility of
deducing the AC effect of a scalar particle from the corresponding
effect for a spinor particle.  In 3+1 dimensions such a connection is
not possible because the anomalous MDM is zero if supersymmetry is an
exact symmetry.  However, in 2+1 dimensions it is possible to have an
anomalous MDM even with exact supersymmetry.

Having demonstrated the relationship between the spinor and the scalar
MDM, we proceed to show that the scalar AC effect is uniquely defined.
We then compute the anomalous MDM at the one loop level, showing how
the scalar form arises in 2+1 dimensions from the coupling of the
scalar to spinors.  This model shows how an AC effect for a scalar
can be generated for non-supersymmetric theories, and we construct
such a model to illustrate the mechanism.
\end{abstract}




The study of topological phases has provided a deep understanding of
quantum systems.  A particularly interesting case of a topological
phase is the Aharonov-Bohm (AB) effect, discovered in 1959 by Aharonov
and Bohm\cite{1}.  The AB effect has been observed
experimentally\cite{2}.  In 1984 Aharonov and Casher
discovered\cite{3} another configuration where a topological phase can
develop, giving rise to what is now called the Aharonov-Casher (AC)
effect.  This effect has also been observed experimentally\cite{4}.
The original AC effect was for a particle with spin and a non-zero
anomalous magnetic dipole moment (MDM) interacting with a two
dimensional electric field perpendicular to the spin polarization
direction.  It was realized that spin 1/2 is a particularly simple and
instructive case, but the AC effect and other related effects have
also been studied for particles with different
spins\cite{5,5a,6,7,8,9,10}.

For a spin-1/2 particle the interaction responsible for the AC effect
is given by the following anomalous MDM interaction,
\begin{eqnarray}
L_m =
-{1\over 2} \mu\bar \psi \sigma^{\mu\nu} \psi F_{\mu\nu}.
\label{amd}
\end{eqnarray}
The topological phase $\theta_{AC}$ developed when the particle
travels along a closed path that encircles a line charge of strength
$\Lambda$ per unit length is $\theta_{AC} = \mu \Lambda$

In general it is possible for a particle with a non-zero anomalous MDM
to have an AC effect.  One would naively think that there should be no
such an effect for a spin-0 particle because there there is no MDM,
anomalous or otherwise.  This naive expectation may not hold in 2+1
dimensions, where anomalous MDM interaction can be interpreted as
an interaction of a current with the dual electric field, as was first
pointed out by Stern\cite{10a}, and noted in the context of the AC
effect for general spins by us\cite{10}.  With this interaction for a
spin-0 particle, an AC like effect can be defined.  This is a very
suggestive way of identifying new topological effects.  However one
still can not be sure that the topological nature of the spin-0 case
has the same origin as the spin-1/2 case.  In this paper we address
the question of whether a spin-0 particle can have an AC effect from
the point of view of supersymmetry.  We find indeed that a spin-0
particle can have an AC effect in 2+1 dimensions.  This will enable us
to discuss the AC effect for a spin zero particle in more general
situations.

We note that Carrington and Kunstatter\cite{10b} followed up Stern's
suggestion and showed that, in Maxwell-Chern-Simons scalar QED in 2+1
dimensions, a magnetic interaction added to the bare Lagrangian leaves
the theory renormalisable at the one loop level.  They also studied
the AC effect of this interaction in the non-relativistic
approximation.  However, they also showed in a later paper \cite{10c},
that at two loop level the scalar theory with a primitive magnetic
interaction is not renormalisable at the two loop level.  This
non-renormalisability is not a difficulty for us, because we use the
MDM interaction added to the Lagrangian as an effective
Lagrangian to demonstrate the supersymmetric connection of the
anomalous MDM of the scalar and the spinor, and never use it to
generate higher order contributions.  We will then discuss two
renormalisable models in which the scalar magnetic moment is generated
at one loop level, just as the Schwinger anomalous magnetic moment of
the electron is generated in QED in $3+1$ dimensions.  These are
\begin{enumerate}
      \item  Supersymmetric
Maxwell-Chern-Simons theory, and
	\item  QED of Yukawa coupled spin 1/2 and
spin 0 particles.
\end{enumerate}

In these models the gauge and the Yukawa couplings have dimensions of
(mass)$^{1/2}$, so both of our models are super-renormalisable, and
have no ultra-violet divergences.  The Chern-Simons masses of the
photon (and photino when present), with the masses of the matter
particles, ensure that the theory is not infra-red divergent
either\cite{10d}.

In this paper we show that the scalar magnetic dipole moment
interaction leads exactly to the AC effect in 2+1 dimensions --- the
non-relativistic approximation is not required.  Carrington and
Kunstatter also observed, in the non-relativistic limit, that the
charged scalar particle with a magnetic interaction can exhibit
anomalous statistics, the scalar equivalent of the result obtained by
us\cite{5a} for Dirac particles in 2+1 dimensions without the
non-relativistic approximation.  Our result on spin-1/2 particles, can
be adapted {\it mutatis mutandis} to apply to the spin-0 interactions
considered here, and shows that the possibility of anomalous
statistics is also an exact result for charged scalar particles with
magnetic interactions in 2+1 dimensions.

In supersymmetry, the spinor (spin-1/2) and the scalar (spin-0)
particles are partners.  If certain effects exist for a spinor, there
should be corresponding effects for the scalar superpartner.
Therefore one would expect that there must be an AC effect for a
spin-0 particle.  This turns out to be not automatically true.  In
fact it was shown some time ago that in 3+1 dimensions, if
supersymmetry is exact, no anomalous MDM can exist for a
spinor\cite{11}.  Of course, in nature supersymmetry is broken, so
there can be a non-zero anomalous MDM for a spinor and therefore an AC
effect.  The corresponding effect for the scalar superpartner need not
exist because supersymmetry is broken.  However this result holds only
in 3+1 dimensions.  In 2+1 dimensions with exact supersymmetry, a spinor can
have an anomalous MDM and therefore the associated AC effect.  Because
supersymmetry is exact, one can uniquely identify the corresponding AC
effect for the scalar superpartner.

In 3+1 dimensions with supersymmetry, a matter field is assigned to a
chiral superfield $\Phi$ which contains the spinor $\psi$ and scalar
$\phi$ as component fields\cite{12}
\begin{eqnarray}
\Phi =
\phi(z) + \sqrt{2} \theta \psi(z) +  \theta \theta
F(z),
\end{eqnarray}
where $z_\mu = x_\mu +i \theta \sigma_\mu \theta$.  $\theta$ is the
anti-symmetric spinor coordinates of the superfield.  $F$ is an
auxiliary field which can be eliminated by the use of the equations of
motion.

The gauge superfield $V$ in the Wess-Zumino gauge is given
by\cite{12}
\begin{eqnarray}
V = -\theta \sigma_\mu \bar \theta A^\mu(x) + i \theta
\theta \bar \theta
\bar \lambda(x) - i \bar \theta \bar \theta \theta
\lambda(x) + {1\over 2}
\theta \theta \bar \theta \theta D(x),
\end{eqnarray}
where $A^\mu$ is the usual gauge field, $\lambda$ is the gaugino field
and $D$ is an auxiliary field.

The usual anomalous MDM interaction of $3+1$ dimensions is contained
in the superfield effective Lagrangian\cite{11}
\begin{eqnarray}
\tilde L_m = ig \Phi^* D_\alpha \Phi W^\alpha,
\end{eqnarray}
where $D_\alpha = \partial/ \partial  \theta^\alpha +
i\sigma^\mu_{\alpha \dot{\alpha}} \bar
\theta^{\dot{\alpha}}\partial_\mu$,
$W^\alpha = (-1/4) \bar
D \bar D D^\alpha V$
with $\bar D_{\dot{\alpha}} =
-\partial/\partial\bar \theta^{\dot{\alpha}} - i
\theta^\alpha
\sigma^\mu_{\alpha\dot{\alpha}}\partial_\mu$.

Expanding $\tilde L_m$ in terms of $\theta$, one would obtain\cite{11}
\begin{eqnarray}
\tilde L_m = E + \chi \theta +
\bar \psi \bar \theta + M \theta \theta + J_\mu \theta
\sigma^\mu \theta
+ \bar \xi \bar \theta \theta \theta.
\end{eqnarray}

The anomalous MDM interaction can be contained only in the $M$ term.
This can be seen easily from a dimensional analysis argument.  The
anomalous MDM interaction is a dimension 5 operator composed of fields
and derivatives.  Inspection of $\tilde L_m$, shows that only the $M$
term is of dimension 5.  However the superfield Lagrangian $\tilde
L_m$ is not a chiral field, so the $M$ term, even being a F-term type,
can not appear in the supersymmetric Lagrangian density.  There is no
way one can obtain a supersymmetric anomalous MDM in 3+1 dimensions.

In 2+1 dimensions the situation changes dramatically.  It is possible
to have a supersymmetric anomalous MDM. In this case the matter field
$\Phi$, the gauge field, and the corresponding $W^\alpha$ are given by
\begin{eqnarray}
&&\Phi(x_\mu, \theta) = \phi(x) + \theta^\lambda
\psi_\lambda(x) -
{1\over 2}\epsilon_{\lambda\tau}\theta^\lambda \theta^\tau
F(x),
\nonumber\\
&&V^\alpha(x_\mu, \theta) = i\theta^\beta (\gamma^\mu
A_\mu(x))^\alpha_\beta - \epsilon_{\lambda\tau}
\theta^\lambda
\theta^\tau \lambda^\alpha(x),\nonumber\\
&&W^\alpha = {1\over 2} D^\beta D^\alpha V_\beta,\;\;\;\;
D_\alpha =
{\partial \over \partial \theta^\alpha} +i \theta^\beta
\epsilon_{\alpha\delta} (\gamma^\mu
\partial_\mu)^\delta_\beta.
\end{eqnarray}
Here $\epsilon_{\alpha\beta}$ is the totally anti-symmetric tensor
with $\epsilon_{12} = 1$.

In 2+1 dimensions, the anomalous MDM interaction has dimension 7/2.
The superfield effective Lagrangian which contains the anomalous MDM
in 2+1 dimensions, given by
\begin{equation}
      \tilde L_m = \Phi^* W_\alpha\Delta^\alpha \Phi,
\end{equation}
(with $\Delta^\alpha = D^\alpha - ie V^\alpha$), has dimension 5/2.
Expanding in terms of the $\theta$, we have
\begin{eqnarray}
\tilde L_m = A + B \theta + C \theta\theta.
\end{eqnarray}
The $C$ term has dimension 7/2 which has the right dimension for the
anomalous MDM interaction and is also supersymmetric.  Therefore in
2+1 dimensions it is possible to have an anomalous MDM interaction and
also the associated AC effect in a supersymmetric theory.  Since the
interaction is supersymmetric, it is possible to identify uniquely the
scalar AC effect.  In the following we provide the detailed
calculation to obtain the form of scalar AC effect interaction.

The supersymmetric Maxwell Lagrangian for the gauge and matter kinetic
energies, and gauge and matter interactions is given by
\begin{eqnarray}
L = \int d^2\theta [{1\over 4}W^\alpha W_\alpha
+ {1\over 2} (\Delta^\alpha \Phi )^* ( \Delta_\alpha \Phi)
- m \Phi^* \Phi ], \label{basic}
\end{eqnarray}
which gives
\begin{eqnarray}
L &=& -{1\over 4} F_{\mu\nu}F^{\mu\nu} + {1\over 2} i \bar
\lambda \gamma^\mu
\partial_\mu \lambda +
\bar \psi i \gamma^\mu D_\mu \psi - m\bar \psi \psi
+ (D^\mu \phi)^*(D_\mu \phi) + L_F\nonumber\\
&+& i e \bar \psi \lambda \phi - ie \bar \lambda \psi
\phi^*.
\label{susy}
\end{eqnarray}
Here $D^\mu = \partial^\mu -ie A^\mu$ is the gauge covariant
derivative.
$L_F$ contains the $F$ term and is given by
\begin{eqnarray}
L_F= F^*F - (m F^* \phi + m \phi^* F).
\end{eqnarray}
Using the equation of motion for $F$,
$F^* = m\phi^*$ and $F = m\phi$ to eliminate the auxiliary
field $F$,
one obtains the scalar mass term
\begin{eqnarray}
L_F = -m^2 \phi^* \phi.
\end{eqnarray}
A straightforward calculation demonstrates that gauge invariant
Chern-Simons mass terms, which have the superfield form
\begin{equation}
     L_{cs} = \frac{M_{cs}}{8}\int d^{2}\theta V^{\alpha}W_{\alpha}
\end{equation}
are generated at one loop level.  Integrating out the Grassman
variables,
\begin{equation}
	L_{cs} =
	+\frac{M_{cs}}{2} \epsilon^{\mu\nu\lambda}
	A_{\mu}\partial_{\nu}A_{\lambda} -
	\frac{M_{cs}}{2}\bar{\lambda}\lambda.
\end{equation}

There is no symmetry principle to forbid the introduction of this $P$
and $T$ violating term, given that we have already introduced a
violation of Parity and Time reversal invariance in the mass term for the
spin-1/2 particles\footnote{See, for example, the discussion of the
discrete symmetry properties of the $2+1$ dimensional fields in
\cite{12a}}.  We therefore allow it to appear in the bare Lagrangian,
which is now referred to as a super-Maxwell-Chern-Simons theory.

A non-zero anomalous MDM interaction
is represented by the introduction of a new superfield term, of the form
$\Phi^*W^\alpha
\Delta_\alpha \Phi$, to the effective Lagrangian.  We will return to
discuss how this term is generated in the perturbation expansion of
the vertex function in the theory, but for the moment we introduce it
into the effective Lagrangian and show how it leads to an AC effect
for the scalar.
\begin{eqnarray}
L_{m} &=& i g \int d^2\theta \Phi^* W^\alpha \Delta_\alpha
\Phi
=
-{1\over 2} g \bar \psi \sigma^{\mu\nu} \psi F_{\mu\nu}
-ig s \epsilon^{\mu\nu\lambda} F_{\mu\nu} \phi^* D_\lambda
\phi\nonumber\\
&+& 2 eg \bar \lambda \lambda \phi^* \phi - g \bar \lambda
\gamma^\mu
\psi (D_\mu \phi)^* - g \bar \psi \gamma^\mu \lambda (D_\mu
\phi)-
ig \bar \psi \lambda F + ig \bar \lambda \psi F^*.
\label{MDML}
\end{eqnarray}
Here $s=\pm$ is defined as $\gamma^\mu\gamma^\nu =
g^{\mu\nu}
+ i s \epsilon^{\mu\nu\lambda}\gamma_\lambda$.

Eliminating the auxiliary field we  obtain the full
Lagrangian
\begin{eqnarray}
L &=& -{1\over 4} F_{\mu\nu}F^{\mu\nu} + {1\over 2} i \bar
\lambda \gamma^\mu
\partial_\mu \lambda\nonumber\\
&+&
\bar \psi i \gamma^\mu D_\mu \psi - m\bar \psi \psi + (D^\mu
\phi)^*(D_\mu \phi)
-m^2\phi^* \phi + i e \bar \psi \lambda \phi - ie \bar
\lambda \psi \phi^*
\nonumber\\
&-& {1\over 2} g \bar \psi \sigma^{\mu\nu} \psi F_{\mu\nu}
-ig s \epsilon^{\mu\nu\lambda} F_{\mu\nu} \phi^* D_\lambda
\phi\nonumber\\
&+& 2 eg \bar \lambda \lambda \phi^* \phi - g \bar \lambda
\gamma^\mu
\psi (D_\mu \phi)^* - g \bar \psi \gamma^\mu \lambda (D_\mu
\phi)\nonumber\\
&-&ig^2\bar \psi \lambda \bar \lambda \psi
- igm \bar \psi \lambda \phi + i gm \bar \lambda
\psi \phi^* 	+\frac{M_{cs}}{2} \epsilon^{\mu\nu\lambda}
	A_{\mu}\partial_{\nu}A_{\lambda} -
	\frac{M_{cs}}{2}\bar{\lambda}\lambda.
\label{full}
\end{eqnarray}

The first term proportional to $g$ in the above equation contains the
usual anomalous MDM term $ \bar \psi \sigma^{\mu\nu} \psi F_{\mu\nu}$
for a spinor with $g = \mu$ compared with Eq.(\ref{amd}).  This term
is responsible for the AC effect of a spin-1/2 particle.  We identify
the second term proportional to $g$ in the above equation to be the
corresponding AC effect term for a scalar particle.

To see the topological nature of these terms
we note that they can be written as
\begin{eqnarray}
L_{AC} = {g\over 2} s F^{\mu\nu} \epsilon_{\mu\nu\lambda}
j^\lambda_F
-{g\over 2} s F^{\mu\nu} \epsilon_{\mu\nu\lambda}
j^\lambda_S,
\label{fullac}
\end{eqnarray}
where $j^\lambda_F = \bar \psi \gamma^\lambda \psi$, and
$j^\lambda_S = i(\phi^*D^\lambda \phi - (D^\lambda \phi^*)
\phi)$ are
the current of the spinor and the scalar, respectively.

   In general the interaction $F^{\mu\nu} \epsilon_{\mu\nu\lambda}
   j^\lambda$ generates a topological phase regardless of the specific
   value of the spin of the particle if the AC conditions required for
   the electric field is satisfied.  This can be seen by studying the
   change of the action $\Delta S$ of the system due to $L_{AC}$, for a
   closed trajectory from time 0 to time T for a point particle with
   velocity $\vec v \propto \vec{j}$.  The topological phase generated
   for the spinor is given by\cite{10}

\begin{eqnarray}
\theta^F_{AC} =- {1\over 2}g \int^T_0F^{\mu\nu}
\epsilon_{\mu\nu\lambda }j^\lambda_F = -g \int^T_0 (\vec S
\cdot \vec
v)dt = -g  \oint \vec S\cdot d\vec r = g \Lambda.
\end{eqnarray}
In the above $S_\mu = (1/2)
\epsilon_{\mu\alpha\beta}F^{\alpha\beta}$.
In the AC electric field configuration, $S_\mu = (0, E_2, -
E_1)$.

Similarly one obtains a topological phase $\theta^S_{AC}$ for the
scalar when the AC conditions are satisfied with $\theta^S_{AC} =
-g\Lambda$.  We note that the topological phases developed for the
spinor and scalar are the same in size and opposite in sign.

Having shown that the term of eq.  (\ref{MDML}), which contains a
supersymmetric anomalous MDM interaction,give an AC effect for the
scalar, we now turn to discuss how such an interaction can be
naturally generated from the well known interactions of eq.
(\ref{basic}) itself.

\begin{figure}[htb]
\centerline{\DESepsf(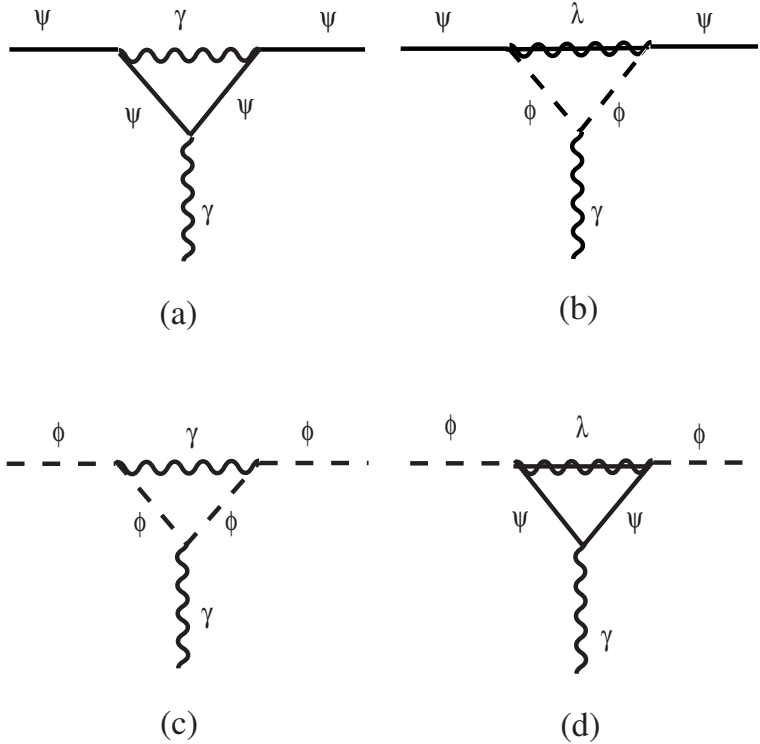 width 10cm)}
\smallskip
\caption{Feynman diagrams for the anomalous MDM.}
\label{figure}
\end{figure}

To this end we study the radiative correction of spinor-photon, and
scalar-photon interactions at the one loop level starting with the
Lagrangian in eq.  (\ref{susy}).  This theory the anomalous MDM,
absent at tree level, can be generated at one loop level, in just the
same way that the Pauli magnetic moment interaction is generated at
one loop level in QED in $3+1$ dimensions.  The relevant diagrams are
shown in Figure 1.  Figure 1.a is the usual QED diagram generating an
anomalous MDM for a spinor.  Due to supersymmetric interactions, there
is an additional diagram, Figure 1.b, contributing to the anomalous
MDM for a spinor.  Evaluating these two diagrams, we obtain an
effective $g$ in eq.  (\ref{full}) with

\begin{eqnarray}
   g = {e^3\over 16\pi m^2} \int^1_0 dx \int^1_x
{y dy \over (y^2-x(y-x)q^2/m^2)^{3/2}},
\label{sg}
\end{eqnarray}
where $q$ is the photon momentum.  Ferrara and Remiddi \cite{11}
showed that in 3+1 dimensions these two contributions canceled, but
we find that in 2+1 dimensions this cancellation does not happen.

The above result is logarithmically infrared divergent when
$q^2$ approaches zero because of the absence of a Chern-Simons mass.  In the
presence of the Chern-Simons mass, the
 theory is free from infrared divergence. In this theory a 
Chern-Simons term will be generated at one loop level from eq. (10). 
The inclusion of these contributions leads to two modifications to 
eq. (19). The first is that the Chern-Simons term, representing a non-zero 
mass $M_{cs}$ for the photon and photino, modifies the denominator of
 eq. (19) 
to $(y^2 + (1-x) M^2_{cs}/m^2 - x(y-x)q^2/m^2)^{3/2}$ and $g$ is infrared 
divergent free. Another modification is that the diagram of Figure 1.a (or of
Figure 1.c) also contributes to the magnetic moment, because of the
term in the propagator proportional to
$M_{cs}\epsilon_{\mu\nu\lambda}q^{\lambda}/q^{2}$ whose contributions 
have been calculated 
by Kogan \cite{13}.  We do not include the Chern-Simons contribution to $g$ in
our calculation, as (unless the bare Chern-Simons mass is
non-vanishing) it is generated at the one loop level in perturbation
theory from the Lagrangian of eq.  (\ref{basic}) and $M_{CS}$ is of
order $e^{2}m$.  Thus its contribution to $g$ is of order $e^{5}m$,
whereas the magnetic moment of eq.  (\ref{sg}) is of order $e^{3}m$.

There are similar radiative corrections to scalar-photon couplings.
These are shown in Figures 1.c and 1.d. Evaluating these diagrams, we
indeed find the second term proportional to $g$ in eq.  (\ref{full})
with the same $g$ as in eq.  (\ref{sg}) generated, as expected.  The
source for a non-zero contribution to $g$ of order $e^3m$
in this case is purely 
Figure 1.d. This is quite different from the spinor case where Figures
1.a and 1.b both contribute to the anomalous MDM at this order.
 As remarked in the
previous paragraph, and reference\cite{13}, a Chern-Simons mass term
gives rise to a scalar anomalous MDM from Figure 1.c, but this
contribution is of order $e^5 m$.

It is clear that the scalar AC effect in the above example is
generated by the Yukawa coupling of a scalar  with two
spinors, Figure 1.d. In 3+1 dimensions, such an interaction can not
generate an AC effect for a scalar.  This shows that the AC effect is
intrinsically a 2+1 dimensional effect.  Supersymmetry then provides
the link between the spinor and scalar AC effects.  With this
understanding one can easily construct a model which generates a
topological AC effect for a scalar in a non-supersymmetric theory by
introducing Yukawa couplings with at least one of the spinor having
non-zero electric charge.  For example, a Yukawa interaction of the
type

\begin{eqnarray}
L_Y = a \bar \psi_1 \psi_2 \phi + H.C.
\end{eqnarray}
will generate a non-zero $g$ through a diagram of the type of Figure 1.d.

\begin{eqnarray}
g &=& {|a|^2 e_1\over 16 \pi^2 m^2_\phi}
\int^1_0 d x \int^1_x {
[(m_1/m_\phi + m_2/m_\phi) y  - m_2/m_\phi] dy \over
[y^2 - x(y-x) q^2/m^2_\phi + m^2_2/m^2_\phi -
y(1-(m^2_1-m^2_2)/m^2_\phi)]^{3/2}}
\nonumber\\
&+& (m_1 \to m_2, m_2\to m_1, e_1 \to e_2),
\end{eqnarray}
where $m_{1,2, \phi}$ are the masses of $\psi_{1,2}$ and $\phi$,
$e_{1,2}$ are the electric charges of $\psi_{1,2}$, respectively.
The above equation reduces to Eq. (\ref{sg})
when $m_1 = m_\phi = m$, $e_1 = e$, and $m_2$ and $e_2$ are both set
to zero, as expected.

We emphasize that, in our models the scalar magnetic moment
interaction is generated by one loop corrections, and does not appear
in the primary Lagrangian, as assumed in the work of Carrington and
Kunstatter\cite{10b}.  We therefore escape the problem that a scalar
magnetic moment in the primary Lagrangian gives non-renormalisable
contributions at two loop level.

To summarize we have studied the supersymmetric AC effect.  In 3+1
dimensions, if supersymmetry is exact, no anomalous MDM interaction
can exist.  But in 2+1 dimensions, we find that a non-zero anomalous
MDM interaction is possible.  The related topological effect for a
scalar is identified which is due to scalar current interaction with
the dual of electric field.  Since the Aharonov-Casher effect is
essentially a phenomenon of two spatial dimensions, we conclude that
there is an Aharonov-Casher effect for a spin-0 particle.  This effect
can be easily generated at one loop level through Yukawa couplings.

\acknowledgments We thank Craig Roberts and Herb Fried for helpful
discussions.
This work was supported in part by the ROC NSC under
grant number NSC 90-2112-M-002-014, by the ROC Ministry of Education
Academic Excellence Project 89-N-FA01-1-4-3, and by the Australian
Research Council.  BHJMcK thanks the Department of Physics at the
National Taiwan University for their hospitality where part of this
work was done.  XGH thanks the School of Physics at the University of
Melbourne for their hospitality while this work was finalized.

\end{document}